\documentclass[english,twocolumn,twocolumn,showpacs,preprintnumbers,amsmath,amssymb,superscriptaddress]{revtex4}
\usepackage[T1]{fontenc}
\usepackage[latin1]{inputenc}
\usepackage{graphicx}

\makeatletter

\makeatletter

\makeatletter

\makeatletter

\makeatletter

\makeatletter

\makeatletter

\makeatletter

\makeatletter

\makeatletter

\makeatletter

\usepackage{color}

\usepackage{dcolumn}

\usepackage{bm}

\newcommand{\pt}{ p_{\rm t}}
\newcommand{\ie}{{\it i.e.}}

 \def\lsim{\mathrel{\rlap{\lower4pt\hbox{\hskip1pt$\sim$}}
    \raise1pt\hbox{$<$}}}         
 \def\gsim{\mathrel{\rlap{\lower4pt\hbox{\hskip1pt$\sim$}}
    \raise1pt\hbox{$>$}}}         

\makeatother

\makeatother

\makeatother

\makeatother

\makeatother

\makeatother

\makeatother

\makeatother

\usepackage{babel}
\makeatother
\begin{document}
\textit{\small QM 2011 contribution} {\small }\textit{\small \vspace{0.5in}
 }{\small \par}

\title{On QGP Formation in pp Collisions at 7~TeV}

\author{Fu-Ming Liu }

\affiliation{Institute of Particle Physics and \rm Key Laboratory of Quark  {\&}  
Lepton Physics (Ministry of Education), Central China Normal University,
Wuhan, China   }

\author{Klaus Werner}

\affiliation{Laboratoire SUBATECH, University of Nantes - IN2P3/CNRS - Ecole des
Mines, Nantes, France }

\date{\today}

\begin{abstract}
The possibility of QGP formation in central pp collisions at ultra-high
collision energy is discussed. Centrality-dependent $\pt$-spectra
and (pseudo)rapidity spectra of thermal photons (charged hadrons)
from pp collisions at 7~TeV are presented (addressed). Minimal-bias
$\pt$-spectrum of direct photons and charged hadrons is compared
under the framework with and without hydrodynamical evolution process. 
\end{abstract}
\maketitle

\section{Introduction}

It is believed that all matter in our universe was in a state called
as the Quark Gluon Plasma (QGP) shortly after the creation of the
universe in the Big Bang. At the mean while, high temperature/density
matter is created in laboratories with high energy accelerators. In
Relativistic Heavy Ion Collider (RHIC), there are many important classes
of experimental observations, such as:

\begin{itemize}
\item Suppression of high $\pt$ hadrons
\item Back-to-back dihadron correlation
\item The scaling of elliptic flow 
\item Direct photon enhancement at low $\pt$
\item ... 
\end{itemize}
They imply QGP formation in central heavy ion collisions at very high energy. The detailed microscopic evolution
procedure of the collision system is not yet fully understood, because
one kind of bricks in the process, the interactions of quarks and
gluons with small momentum exchange, $\ie$, below $\Lambda_{QCD}$,
is barely known. Nevertheless, direct photons are regarded as an ideal
probe of the evolution procedure of the collision system due to two
reasons:

\begin{enumerate}
\item They can be produced during the whole evolution via the interaction
(soft and hard) between quarks and gluons, and the interaction between
hadrons. 
\item They can penetrate the collision system without further interaction
because of their big mean free path. 
\end{enumerate}
By definition, one can imagine the creation of a QGP in proton-proton
collisions at ultra-high energies from the intuition. The problems
are mainly two aspects:

\begin{enumerate}
\item How to construct a reliable calculation in order to learn more; 
\item What are the sensitive signals for QGP formation in pp collisions. 
\end{enumerate}
In the following sections, we will focus on the two questions, especially
on soft physics, $\ie$, the production of the most copious particles.

\section{How to construct the calculation?}

First of all, we have to classify the centrality. In heavy ion collisions,
this is determined by how many pairs of nucleon collisions, measured
via the multiplicity or the transverse energy (at midrapidity region).
In pp collisions, centrality looks like a strange word. But centrality
becomes important in pp collisions with the increase of the collision
energy $\sqrt{s}$ because partons at low-$x$ in proton becomes more
and more active. We will see the details in the following.

In AA collisions, pairs of nucleons from projectile and target interact
in parallel at early stage, followed by many secondary interactions.
Now we are treating the early stage of pp collisions, where pairs
of partons from projectile proton and target proton interact in parallel.
While the former one is treated with Glauber model, the latter one
can be treated with Gribov theory~\cite{Gribov}. Two examples of
centrality in pp collisions are shown in Fig.1. Example (a): only
two quark-diquark strings are excited between projectile proton and
target proton to produce particles. Example (b): two gluons are emitted
additionally, one from projectile proton and the other from target
protons, and convert into a quark and an antiquark at each side, then
two pairs of quark-antiquark interaction between projectile and target
are formed. More pairs of partonic interaction can occur through more
gluon emissions, which provide more central pp collisions.

\begin{figure}
\includegraphics[scale=0.7]{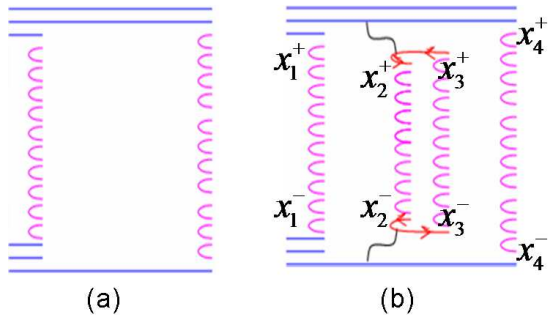}

\caption{\label{1} (Color Online) Two examples of centrality in pp collisions.
For details see the text. Energy momentum fractions $x_{i}^{+}$ and
$x_{i}^{-}$ in (b) are labeled. Picture comes from ~\cite{Liu:2003wja}.}
\end{figure}

How is the centrality study relevant to collision energy $\sqrt{s}$?
In Fig.1(b) energy momentum fractions $x$ carried by partons from
protons are labeled for more convenient discussion. A chain of quark
and antiquark (or diquark) between projectile and target is called
as a string. The string has a total energy $x^{+}x^{-}\sqrt{s}$,
where $x^{+}$ and $x^{-}$ are the energy momentum fractions of the
partons at projectile and target side, respectively. Partons at two
ends of strings will fly away according to their energy and momentum.
A string will break when its two ends becomes further and further,
because the $q\bar{q}$ linear potential should not exceed the total
energy of the string. The fragments of strings are the hadrons produced.
Partons at low-$x$ in proton becomes more active with the increase
of the collision energy $\sqrt{s}$ , because the energy threshold
for strings to produce particles remains the same:

\[
x^{+}x^{-}\sqrt{s}>E_{0},\]
 where $E_{0}$ is an energy scale around 1~GeV. Thus, an ultra-high
collision energy $\sqrt{s}$ means partons at very low $x$ are energetic
enough to go out the Dirac sea and involve particle production. Because
the density of partons in a proton increases rapidly with the decrease
of $x$, many low-$x$ partons may form strings to produce particles.
So at ultra-high collision energy, the number of strings may fluctuate
largely from event to event, and the classification of centrality
becomes important.

Let's make a comparison of centrality between AA collisions and pp
collisions to see more details:

\begin{enumerate}
\item In the case of AA, the phase space distribution of nucleons in a heavy
nuclei is much simpler. The spatial distribution is the widely tested
Wood-Saxon type. The total number of nucleons in a heavy nuclei is
the mass number $A$ and the total energy is almost equally shared
by them. In the case of pp, each sampling of parton energy from a
proton should obey the parton distribution function in proton (PDF).
However, the PDF at low-$x$ is much less explored. There are also
constraints from conservation law, $\ie$, energy conservation requires
$\sum_{i}x_{i}=1$ at both projectile and target sides. 
\item In the case of AA, the reference of pp collisions can be fully studied
directly, concerning to all kind of observables. In the case of pp,
the reference of partonic collisions can only be obtained indirectly,
$\ie$, via $e^{+}e^{-}\rightarrow q\bar{q}$ process, due to quark
confinement. The main parameters of string fragmentation have been
determined phenomenologically. 
\item In the case of (central) AA collisions, many particles are produced
from the collisions of nucleon pairs, then secondary interaction between
produced particles cannot be ignored. A large number of secondary
interaction in the collision system can be treated macroscopically
with hydrodynamics. In the case of (central) pp collisions at very
high energy, many particles are produced from the collisions of many
parton pairs, too. The secondary interaction between produced particles
cannot be ignored, either. We also take hydrodynamics to treat the
evolution of the many-body system. 
\end{enumerate}
So we construct the calculation like this:

\begin{enumerate}
\item Initial condition~\cite{Drescher:2000ec}: For each given centrality
of pp collisions, the initial energy momentum tensor is obtained from
strings or string fragments via: \[
T_{{\rm hydro}}^{\mu\nu}(\tau_{0},\vec{r})=T_{{\rm particle}}^{\mu\nu}(\tau_{0},\vec{r})\equiv\frac{\sum_{i\in\Delta V}p_{i}^{\mu}p_{i}^{\nu}}{\Delta V}.\]

\item Evolution: The evolution of energy momentum tensor is governed by
the conservation law, \[
\partial_{\mu}T^{\mu\nu}=0,\]
 solved in full 3+1D space-time $(\tau,x,y,z)$ with equation-of-state
taken from Lattice results, then decomposed as the following terms:
\[
T^{\mu\nu}=(e+P)u^{\mu}u^{\nu}-Pg^{\mu\nu}+\Pi^{\mu\nu},\]
 to obtain energy density $e$, pressure $P$, local four flow velocity
$u^{\mu}$, where $g^{\mu\nu}={\rm diag}(1,-1,-1,-1)$ is the metric
tensor and the viscosity term $\Pi^{\mu\nu}$ is set to zero in ideal
hydrodynamics. 
\item Freeze-out: We take the same freeze-out condition as in heavy ion
collisions, $\ie$, $e^{th}=0.08{\rm GeV}/{\rm fm}^{3}$ or $T^{th}=100\,{\rm MeV}$.
Thus soft hadron production can be obtained with the Cooper-Fry formula. 
\end{enumerate}
Thermal photons can be produced during the whole evolution procedure,
calculated via photon emission rate with the above obtained local
temperature and flow velocity. Thermal photons are the main source
of direct photons at low transverse momentum region, as we know from
heavy ion collisions, similar to the soft hadron production from the
bulk.

One may argue that the pp collision system is too small to employ
hydrodynamics. This is first a question of collision energy and collision
centrality. At low collision energies, the pp collision system has
a small size. But at ultra-high collision energies, there is a big
fluctuation of centrality, as we discussed above. Thus, for the central
pp collisions at very high energy, a relatively big size may be formed,
$\ie$, implied by the Bose-Einstein correlation or HBT study.

At the other hand, a high energy density region must be formed when
the two highly compressed protons are overlapped. Then many constituent
particles in the system are partons due to the high energy density,
whose mean free path may be very small, because $\lambda^{-1}=\rho\sigma$
and

\begin{enumerate}
\item the cross section of soft partonic interaction $\sigma$ is not precisely
known, but quite big due to strong coupling at the non-perturbative
region; 
\item The parton density $\rho$ is certainly very high at high energy density. 
\end{enumerate}
Thus the employment of hydrodynamics might be legal. Anyway, our knowledge
on the properties of the created quark matter is very limited.

\section{Centrality Dependent Results and Minimal-Bias Results}

In the study of heavy ion collisions, we have a good reference, which
is the nucleon-nucleon (or pp ) collisions. A typical quantity is
so called nuclear modification factor, of any kind of identified particles.
However, in our present case of pp collisions, this kind of measurement
are not available. However, similar to the ratio between central collisions
and peripheral collisions $R_{cp}$, special hints can be found in
centrality dependent results, $\ie$, Ridge behavior observation in
high multiplicity pp events~\cite{cms_ridge}. To understand those
results, let's first check something basic.

\begin{figure}
\includegraphics[scale=0.8]{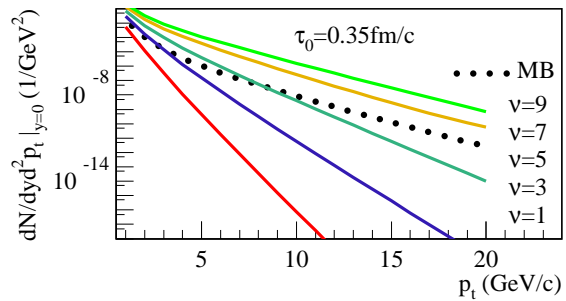}

\caption{\label{2} (Color Online) $\pt$ spectra of thermal photons from
pp collisions at centrality $\nu=1,3,5,7,9$ for solid lines (from
down to up). The dotted line gives minimal-bias result.}
\end{figure}

\begin{figure}
\includegraphics[scale=0.8]{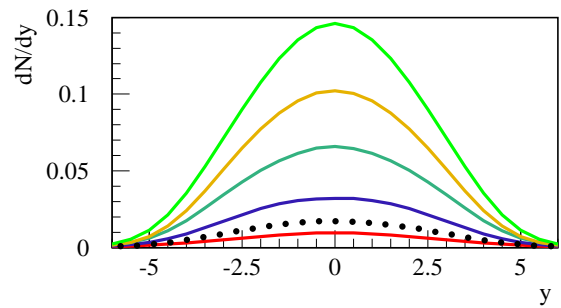}

\caption{\label{3} (Color Online) Rapidity spectra of thermal photons from
pp collisions at centrality. Same notation as Fig.~\ref{2}.}
\end{figure}

In Fig.~\ref{2} and Fig.~\ref{3}, centrality dependent pt-distribution
and rapidity distribution of thermal photons are shown. The centrality
is labeled as $\nu$ which is the number of Pomerons (a pair of strings
is called as a Pomeron). The result is easy to understand: For more
central events, more energy is deposited at the reaction region (less
energy is brought away by leading particles). Thus more secondary
collisions at the reaction region to emit more thermal photons. The
same type results on charged hadrons are not shown, but the centrality
dependence is similar. We simply parameterize the obtained pseudorapidity
density of charged hadrons $dn/d\eta|_{\eta=0}$ as a function of
centrality $\nu$ :\[
\frac{dn}{d\eta}|_{\eta=0}=2.8147\nu+4.3477.\]

\begin{figure}
\includegraphics[scale=0.4]{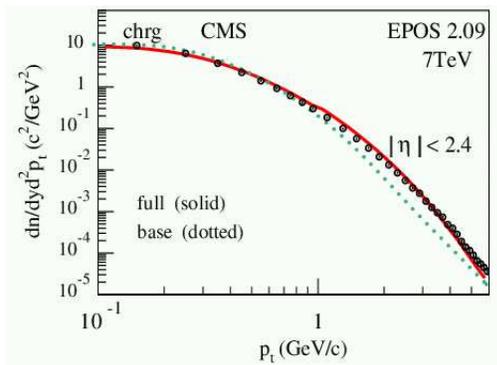}

\caption{\label{4} (Color Online) Minimal-bias $\pt$ spectrum of charged
hadrons from pp collisions at 7~TeV, without evolution procedure
(dotted line) and with hydrodynamical evolution (solid line) are compared
with CMS data points ~\cite{Khachatryan:2010us}. Picture comes
from ~\cite{Werner:2010ss}. }
\end{figure}

What is the consequence after considering the evolution procedure
in pp collisions? Let's first review the minimal-bias transverse momentum
spectrum of both charged hadrons~\cite{Werner:2010ss} and direct
photons~\cite{Liu:2008eh}. In Fig.~\ref{4} the transverse momentum
spectrum of charged hadrons from pp collisions at 7~TeV without evolution
procedure (dotted line) and with hydrodynamical evolution procedure
(solid line) is shown together with the data points from CMS~\cite{Khachatryan:2010us}.
We can see that the evolution procedure (the collective flow) has
very little effect in charged hadron production. Nevertheless, it
is the reason why the Ridge is observed in high multiplicity pp events
~\cite{Werner:2010ss}.

\begin{figure}
\includegraphics[scale=0.7]{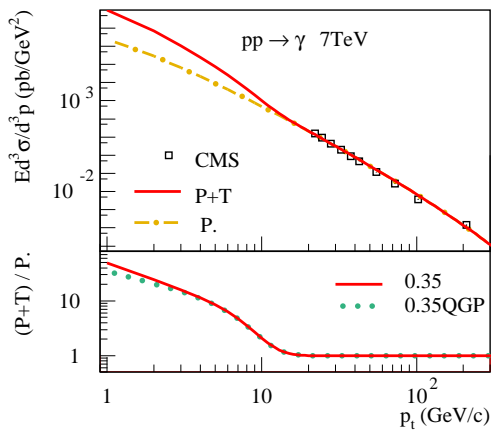}

\caption{\label{5} (Color Online) Upper panel: minimal-bias $\pt$ spectrum
of direct photons from pp collisions at 7TeV without evolution procedure
(dashed-dotted line) and with evolution procedure (solid line) are
compared with CMS data~\cite{Khachatryan:2010fm}. Lower panel:
the ratio of the two lines in upper panel. See text for more details.}
\end{figure}

Direct photons are much more sensitive to this hydrodynamical evolution
procedure, because a completely new source, thermal photons (T), will
be added to the conventional prompt photons (P). In Fig.~\ref{5}
the upper panel, the minimal-bias $\pt$ spectrum of direct photons
from pp collisions at 7~TeV is shown in both cases, with hydrodynamical
evolution procedure (solid line) and without (dashed-dotted line).
The CMS data points~\cite{Khachatryan:2010fm} are also shown, but
not at the sensitive region. The ratio of the two lines are also plotted
as solid line in the lower panel. To prove the QGP formation in pp
collision, we count only thermal photons emitted from purely QGP phase
in the evolution procedure and ignore thermal photon emitted from
hadronic gas. The result is plotted in the lower panel as dotted line.
We can see that the thermal photons are mostly from the pure QGP phase.
This is a direct theoretical proof for QGP formation in pp collisions~\cite{Liu:2011dk}.%

\section{Conclusion and Discussion}

Centrality becomes important in pp collisions at ultra-high collision
energy, because partons at low-$x$ in the proton become more active
in particle production. The centrality-dependent $\pt$ spectrum and
(pseudo)rapidity spectrum of both thermal photons and charged hadrons
show a similar centrality-dependence in heavy ion collisions~\cite{Back:2001bq}.
Whereas the evolution procedure effects very little in the minimal-bias
$\pt$ spectrum and rapidity spectrum of charged hadrons, it has a
dramatical influence on those of direct photons, because a completely
new source, thermal photons, will be added to prompt photon production
at the low $\pt$ region. Additionally, the thermal photons from pure
QGP phase is so much pronounced, that it can be proposed as a direct
signal of QGP formation in pp collisions.

Although direct photons are heavily polluted by decay photons at low
$\pt$ region, special technique has been developed by PHENIX~\cite{PHENIX10}
to treat this challenge, which did meet the prediction well~\cite{Liu:2008eh}.

\begin{acknowledgments}
This work is supported by the Natural Science Foundation of China
under the project No. 10975059. 
\end{acknowledgments}

\end{document}